\documentclass[x11names]{llncs}
\usepackage[utf8]{inputenc}
\usepackage[OT1]{fontenc}
\usepackage{amsfonts}
\usepackage{listings}
\usepackage{courier}
\usepackage{pgfplotstable}
\usepackage{times}
\usepackage{xspace}
\usepackage{tikz}
\usepackage{amssymb}
\usepackage{amsmath}
\usepackage{comment}
\usepackage{stmaryrd}
\usepackage{subcaption}
\usepackage{multirow}
\usepackage{wrapfig}
\usepackage{pifont}
\usepackage{bigdelim}
\usepackage{mathtools}
\usepackage{letltxmacro}
\usepackage{courier}
\usepackage{hyperref}
\usepackage{syntax}
\usepackage{longtable}
\usepackage{pdflscape}

\usetikzlibrary{shapes,arrows,chains,fit,backgrounds}

\tikzstyle{bef}=[align=left,font=\ttfamily,node distance=3mm,text width=\textwidth*.4]
\tikzstyle{aft}=[align=left,font=\ttfamily,node distance=3mm,text width=\textwidth*.5]
\tikzstyle{acsl}=[inner sep=0mm,fill=blue!15]
\tikzstyle{genc}=[inner sep=0mm,fill=blue!15]

\tikzstyle{common}=[node distance=15mm,align=center]
\tikzstyle{test}=[draw,trapezium,trapezium left angle=70,
  trapezium right angle=-70,common]
\tikzstyle{op}=[draw,common]
\tikzstyle{data}=[draw,common,ellipse,node distance=35mm]
\tikzstyle{ce}=[data]
\tikzstyle{arrow}=[draw,->]
\tikzstyle{darrow}=[draw,o->]

\newcommand{\cmark}{\ding{51}\xspace}%
\newcommand{\xmark}{\ding{55}\xspace}%
\newcommand{\cko}{--\xspace}%
\newcommand{\ctimeout}{\ding{36}\xspace}%
\newcommand{\ctodo}{\ding{48}\xspace}%
\renewcommand{\phi}{\varphi}

\newcommand{\framac}{\textsc{Frama-C}\xspace}
\newcommand{\pathcrawler}{\textsc{Path\-Craw\-ler}\xspace}
\newcommand{\acsl}{\textsc{ACSL}\xspace}
\newcommand{\eacsl}{\textsc{E-ACSL}\xspace}
\newcommand{\Wp}{\textsc{Wp}\xspace}
\newcommand{\stady}{\textsc{StaDy}\xspace}

\newcommand{\toolname}{\textsc{RPP}\xspace}

\lstdefinelanguage{pretty-ACSL}{%
  escapechar={},
  literate=
   {==}{{$==$}}2
   {==>}{{$\Rightarrow$}}1
   {integer\ i}{{i$\,\in \mathbb{Z}\,$}}4
   {integer\ j}{{j$\,\in \mathbb{Z}\,$}}4
   {integer\ k}{{k$\,\in \mathbb{Z}\,$}}4
   {integer\ m}{{m$\,\in \mathbb{Z}\,$}}4
   {integer\ l}{{l$\,\in \mathbb{Z}\,$}}4
   {\\forall}{{$\forall$}}1
   {\\exists}{{$\exists$}}1
   {integer}{{$\mathbb{Z}$}}1
   {real}{{$\mathbb{R}$}}1
   {&&}{{$\wedge$}}1
   {||}{{$\vee$}}1
   {!=}{{$\neq$}}1
   {<}{{$<$}}1
   {<=}{{$\le~$}}1
   {>}{{$>$}}1
   {>=}{{$\ge~$}}1
   {<==>}{{$\Leftrightarrow$}}1,
  morekeywords={assert,assigns,assumes,axiom,axiomatic,behavior,behaviors,
    boolean,breaks,complete,continues,data,decreases,disjoint,ensures,
    exit_behavior,ghost,global,inductive,invariant,lemma,logic,loop,
    model,predicate,relational,reads,requires,sizeof,strong,struct,terminates,
    \\callpure,\\callresult,\\callset,\\call,\\from,
    type,union,variant,uchar,byte,typically,\\result,\\old,\\at,\\valid,
    \\separated,\\nothing,Pre,Post,Here,\\sum,\\numof,\\call,\\from},
  alsoletter={\\},
  morecomment=[l]{//}
}
\lstnewenvironment{pretty-codeACSL}{\lstset{language=pretty-ACSL,stepnumber=0}}{\smallskip}

\lstdefinelanguage{ACSL}{%
  escapechar={},
  literate=,
  morekeywords={assert,assigns,assumes,axiom,axiomatic,behavior,behaviors,
    boolean,breaks,complete,continues,data,decreases,disjoint,ensures,
    exit_behavior,ghost,global,inductive,invariant,lemma,logic,loop,
    model,relational,predicate,reads,requires,sizeof,strong,struct,terminates,
    \\callpure,\\callresult,\\callset,\\call,\\from,
    type,union,variant,uchar,byte,typically,\\result,\\old,\\at,\\valid,
    \\separated,\\nothing,Pre,Here,Post,\\exists,\\forall,\\sum,\\numof},
  alsoletter={\\},
  morecomment=[l]{//}
}
\lstnewenvironment{codeACSL}{\lstset{language=ACSL,stepnumber=0}}{\smallskip}

\lstdefinestyle{pretty-c}{language={[ANSI]C},%
  alsolanguage=pretty-ACSL,%
  moredelim={*[l]{//}},%
  deletecomment={[s]{/*}{*/}},
  moredelim={*[l]{//@}},%
}

\lstdefinestyle{c}{language={[ANSI]C},%
  alsolanguage=ACSL,%
  moredelim={*[l]{//}},%
  deletecomment={[s]{/*}{*/}},
  moredelim={*[l]{//@}},%
}

\newcolumntype{C}[1]{>{\centering\let\newline\\\arraybackslash\hspace{0pt}}m{#1}}

\begin{document}

% Added for page numbers
%\pagestyle{headings}

\title{
Static and Dynamic Verification of Relational Properties on Self-Composed C Code 
%Self-composition to Prove Relational Properties in Annotated C Program
}
%% \author{
%%   \mbox{}
%%   \hspace{-4mm}
%%   Lionel Blatter\inst{1,2} \and
%%   Nikolai Kosmatov\inst{1} \and
%%   Pascale Le Gall\inst{2}  \and
%%   Virgile Prevosto\inst{1} \and
%%   Guillaume Petiot\inst{1}
%% }
%% \institute{
%%   CEA, List, Software Reliability and Security Laboratory,
%%   91191 Gif-sur-Yvette France\\
%%   \email{firstname.lastname@cea.fr}
%%   \and
%%   Laboratoire de Mathématiques et Informatique pour la Complexité et les Systèmes\\
%%   CentraleSupélec, Université Paris-Saclay, 91190 Gif-sur-Yvette France\\
%%   \email{firstname.lastname@centralesupelec.fr}
%% }

\author{
  \mbox{}
  \hspace{-4mm}
Lionel Blatter\inst{1,2} \and
Nikolai Kosmatov\inst{1} \and
Pascale Le Gall\inst{2} \and \\
Virgile Prevosto\inst{1} \and
Guillaume Petiot\inst{1}}
\institute{
CEA, List, Software Reliability and Security Lab,  PC 174, 91191 Gif-sur-Yvette France\\
\email{firstname.lastname@cea.fr}
\and
CentraleSupelec, Université Paris-Saclay, 91190 Gif-sur-Yvette France\\
\email{firstname.lastname@centralesupelec.fr}}

%\date{\today}
\maketitle

\begin{abstract}
%% Deductive verification provides a powerful tool to show functional properties
%% of a given program.
%% %Commonly, these properties relate to a single program.
%% However, in practice, many properties of interest link several program calls.
%% This is for instance the case for non-interference, continuity and monotony.
%% Other examples relate sequences of function calls, for instance to
%% show that decrypting an encrypted message with the appropriate key gives back
%% the original one message.
%% Such properties cannot be expressed directly in the traditional setting
%% used by modular deductive verification, but are amenable to verification
%% %formal techniques: Cartesian Hoare Logic,
%% %Relational Hoare Logic, Self-Composition or Program Product.
%% %This  paper will focus on 
%% through \emph{self-composition}.
%% %and its
%% %application in the verification of relational properties in C programs.
%% This paper presents a verification tool dedicated to relational properties,
%% in the form of a Frama-C plug-in called \toolname and based on self-composition.
%% It supports functions with side effects and recursive functions.
%% Our initial experiments on existing benchmarks confirm that \toolname 
%% is useful to prove relational properties.
%% %We give illustrative examples and present the application of the tool
%% %in the particular case of the frame rule verification.

Function contracts are a well-established way of formally specifying the
intended behavior of a function. However, they usually only describe what
should happen during a single call. Relational properties, on the other hand,
link several function calls. They include such properties as non-interference,
continuity and monotonicity. Other examples relate sequences of function calls, for
instance, to show that decrypting an encrypted message with the appropriate key
gives back the original message. Such properties cannot be expressed
directly in the traditional setting of modular deductive verification, but are
amenable to verification through \emph{self-composition}. This paper presents a
verification technique dedicated to relational properties in C programs and its implementation in the form of a \framac plugin
called \toolname and based on self-composition. It supports functions with side effects
and recursive functions. The proposed approach makes it possible to prove a relational property,
to check it at runtime, to generate a counterexample using testing and to use it as a hypothesis in the subsequent verification.
Our initial experiments on existing benchmarks confirm
that the proposed technique is helpful for static and dynamic analysis of relational properties.

\end{abstract}

%-------------------------------------------------------------
\section{Introduction}

\label{sec:intro}
\paragraph{Context.}
Deductive verification techniques provide powerful methods for formal verification 
of properties expressed in Hoare Logic~\cite{Floyd1967,Hoare1969}.
In this formalization, also known as axiomatic semantics, a program is seen as a
predicate transformer, where each instruction $S$ executed on a state verifying
a property $P$ leads to a state verifying another property $Q$.
This is summarized in the form of \emph{Hoare triples} as $\{P\}S\{Q\}$.
In this setting, $P$ and $Q$ refer
to states before and after a single execution of a program $S$. It is possible
in $Q$ to refer to the initial state of the program, for instance to specify
that $S$ has increased the value stored in variable \texttt{x}, but one
cannot express properties that refer to two distinct executions of $S$,
even less
properties relating executions of different programs $S_1$ and $S_2$.
As will be seen in the next sections, such properties, that
we will call \emph{relational properties} in this paper,
occur quite regularly in
practice. Hence, it is desirable to provide an easy way to specify them and
to verify that implementations are conforming to such specification.
A simple example of a relational property is monotonicity of a function $f$:
$x<y \Rightarrow \mathtt{f}(x) < \mathtt{f}(y)$.

Several theories and techniques exist for handling relational properties.
First, Relational Hoare Logic~\cite{DBLP:conf/popl/Benton04} is
mainly used to show the correctness of program transformations, \emph{i.e.} the
fact that the result of the transformation
preserves the original semantics of the code.
Then, Cartesian Hoare Logic~\cite{SousaD16} allows for the verification of
$k$-safety properties, that is, properties over $k$ calls of a function.
The \textsc{Descartes} tool is based on Cartesian Hoare Logic and has been used
to verify anti-symmetry, transitivity and extensionality of various comparison
functions written in Java.
A decomposition technique using abstract interpretation
is presented in \cite{DBLP:conf/pldi/AntonopoulosGHK17}
for verification of $k$-safety properties.
The method is implemented in a tool called \textsc{Blazer} and
used for verification of non-interference and absence of timing
channel attacks.
A relational program reasoning based on an intermediate program representation
in LLVM is proposed by \cite{DBLP:journals/jar/KieferKU18}. The method supports loops
and recursive functions and is used for checking program
equivalence.
Finally, self-composition~\cite{DBLP:journals/mscs/BartheDR11} and its
refinement Program Products~\cite{prograprodu} propose
theoretical approaches to prove relational properties by reducing the
verification of relational properties to a standard deductive verification
problem.

\paragraph{Motivation.}
In the context of the \acsl specification language~\cite{ACSL} and the deductive 
verification plugin \Wp of \framac \cite{Frama-C}, the necessity to deal with 
relational properties has been faced in various verification projects.
For example, we can extract the following quote from
a work on verification of continuous monotonic functions
in an industrial case study on smart sensor software~\cite{BishopBC13}
(emphasis ours):
\begin{quote}
After reviewing around twenty possible code analysis tools, 
we decided to use \framac, which fulfilled all our 
requirements \emph{(apart from the specifications involving the 
comparison of function calls)}.
\end{quote}
The authors attempt to prove the monotonicity of some functions
(i.e., if $x \leq y$ then $f(x) \leq f(y)$) using 
\framac/\Wp plugin.
To address the absence of support for relational properties in \acsl and \Wp,
they perform a manual transformation \cite{BishopBC13} 
consisting in writing an additional function simulating the call 
to the related functions in the property. Broadly speaking, this amounts to
manually perform self-composition. This technique is indeed quite simple and
expressive enough to be used on many relational properties. However, applying
it manually is relatively tedious, error-prone, and does not provide a completely
automated link
between three key components: \textsl{(i)}~the specification of the property,
\textsl{(ii)}~the proof that the implementation satisfies the property,
and \textsl{(iii)}~the ability to use the property as hypothesis
in other proofs (of relational as well as non-relational properties).
Thus, the lack of support for relational properties can be a major obstacle to
a wider application of deductive verification in academic and industrial
projects.
Finally, another motivation of this work was to obtain a solution compatible
with other techniques than deductive verification, notably dynamic analysis.

%% In the context of library specification, relational properties can help specify
%% higher order properties. 
%% The properties will be specified, and used as hypothesis but not proved.
%% The PISCO project\footnote{\url{http://www.systematic-paris-region.org/en/projets/pisco}}, an industrial case study on verification of
%% software using hardware-provided cryptographic primitives (PKCS\#11 standard)
%% required tying together different functions with properties such as
%% \smash{$\mathrm{Decrypt}(\mathrm{Encrypt}(Msg,PrivKey),PubKey)=Msg$}.

%% Other examples include properties of data structures,
%% such as matrix transformations (e.g. $(A+B)^\intercal = A^\intercal+B^\intercal$ or
%% $\mathrm{det}(A) = \mathrm{det}(A^\intercal)$), 
%% the specification of
%% $\mathrm{Push}$ and $\mathrm{Pop}$ over a stack~\cite{ACSLbyExample},
%% or parallel program specification 
%% (e.g., 
%% \smash{$\mathrm{map}(\mathrm{append}(l_1,l_2)) = \mathrm{append}(\mathrm{map}(l_1),\mathrm{map}(l_2))$}
%% in the MapReduce approach.

 \paragraph{Contributions.}
To address the absence of support for expressing relational properties
in \acsl and for verifying such properties in the \framac platform,
we implemented a new plugin called \toolname.
This plugin allows the specification and verification of properties
invoking any (finite) number of calls of possibly dissimilar functions
with possibly nested calls, and to use the proved properties as hypotheses
in other proofs.
A preliminary version of \toolname has been described
in a previous short paper~\cite{DBLP:conf/tacas/BlatterKGP17}.
 However, it suffered from major limitations.
Notably, it could only handle pure, side-effect free functions, which in
the context of the C programming language is an extremely severe constraint.
Similarly, the original syntax to express relational properties is not expressive enough
and requires some additional constructs, in order to properly specify
relational properties of functions with side-effects.
The previous work~\cite{DBLP:conf/tacas/BlatterKGP17} did not address dynamic analysis of
relational properties either.

The current paper will thus focus on the
extensions that have been made to the original \toolname design
and implementation, as well as its evaluation. Its main contributions include:
\begin{itemize}
\item a new syntax for relational properties;
\item handling of side effects;
\item handling of recursive functions;
\item evaluation of the approach over a suitable set of illustrative examples;
\item experiments with runtime checking of relational properties
  and counterexample generation when a property cannot be
  proved in the context of \toolname.
\end{itemize}

\paragraph{Outline.}
The remainder of this paper is organized as follows. First,
in Section~\ref{subsec:rpp} we briefly recall
the general idea of relational property verification with
\toolname in the case of pure functions using self-composition.
Then, in Section~\ref{subsec:sideeffect},
we show how to extend this technique to the verification of
relational properties over functions with
side effects (access to global variables and pointer dereference). 
Another extension, described in
Section~\ref{subsec:recursive} allows considering recursive functions.
We demonstrate the capacities of \toolname 
by using it on the adaptation
to C of the benchmark proposed for Java in~\cite{SousaD16} 
and our own set of test examples (Section~\ref{subsec:example}).
Finally, we show in Section~\ref{subsec:test}
that \toolname can also be used to check relational properties
at runtime and/or to generate a counterexample using testing,
and conclude in Section~\ref{sec:conclusion}.

% Local Variables:
% compile-command: "pdflatex main.tex"
% mode: latex
% TeX-master: main.tex
% TeX-PDF-mode: t
% mode: flyspell
% ispell-local-dictionary: "american"
% End:

%----------------------------
\section{Context and Main Principles}
\label{subsec:rpp}
%--------------------------------------------------------1---------------
\toolname (Relational Property Prover) is a solution designed and implemented
as a plugin of \framac~\cite{Frama-C}, an extensible framework
dedicated to the analysis of C programs.
\framac offers a specification language, called \acsl~\cite{ACSL}, and a deductive verification plugin, \Wp~\cite{WP}, 
that allow the user to specify the desired program properties as function contracts 
and to prove them.  
A typical \acsl function contract may include a precondition 
(\lstinline'requires' clause stating a property that must hold each time the function is called)
and a postcondition  (\lstinline'ensures' clause that must hold
when the function returns),
as well as a frame rule (\lstinline'assigns' clause indicating which parts
 of the global program state the function
is allowed to modify). \lstinline'assigns' clauses may be refined by
\lstinline'\from' directives, indicating for each memory location $l$
potentially modified by the function the list of memory locations that are read
in order to compute the new value of $l$.
Finally, an assertion (\lstinline{assert} clause) can also specify a local
property at any function statement.

\Wp is based on Hoare logic
and generates Proof Obligations (POs) using Weakest
Precondition calculus: given a property $Q$ and a fragment of code $S$,
it is possible to compute the minimal (weakest)
condition $P$ such that $\{P\}S\{Q\}$ is a valid Hoare triple.
When $S$ is the body
of a function $f$, POs are
formulas expressing that the precondition of $f$ implies the
weakest condition necessary for the postcondition (or assertion) to hold
after executing $S$. POs can then be discharged either automatically
by automated theorem provers (e.g.
Alt-Ergo, %\footnote{\url{https://alt-ergo.ocamlpro.com}},
CVC4, %\footnote{\url{http://cvc4.cs.nyu.edu}},
Z3\footnote{See, resp., \url{https://alt-ergo.ocamlpro.com},
\url{http://cvc4.cs.nyu.edu},
\url{https://z3.codeplex.com/}})
or with some help from the user {\it via} a
proof assistant (e.g. Coq\footnote{See \url{http://coq.inria.fr/}}).

\framac also offers an executable subset of \acsl, called \eacsl~\cite{Delahaye/SAC13,E-ACSL},
that can be transformed into executable C code. It is thus
compatible with dynamic analysis,
such as runtime assertion checking of annotations
using the \eacsl plugin~\cite{Delahaye/SAC13,VorobyovKS-e-acsl-segmodel}
or with counterexample generation (in case of a proof failure)
using the \stady plugin~\cite{Petiot/SCAM14,DBLP:conf/tap/PetiotKBGJ16}.

Function contracts allow specifying the behavior of a single
function call, that is, properties of the form
``If $P(s)$ is verified when calling $f$ in state $s$, $Q(s')$ will be verified
when $f$ returns with state $s'$''.
However, it is not possible to specify \emph{relational properties}, that
relate several function calls. Examples of such properties include monotonicity
($x<y \Rightarrow \mathtt{f}(x) < \mathtt{f}(y)$), anti-symmetry
($\mathtt{compare}(x,y)=-\mathtt{compare}(y,x)$) or
transitivity
($\mathtt{compare}(x,y)\leq 0 \wedge \mathtt{compare}(y,z) \leq 0 \Rightarrow
 \mathtt{compare}(x,z)\leq 0$). \toolname addresses this issue by providing
an extension to \acsl for expressing such properties and a way to prove them.
More specifically, \toolname works like a preprocessor for \Wp: given a
relational property and the definition of the C function(s) involved in the
property, it generates a new function together with plain \acsl annotations
whose proof (using the standard \Wp process)
implies that the relational property holds for the original code.
As we show below, this encoding of a relational property
is also compatible with dynamic analysis (runtime verification or counterexample generation).

%------------------------------------------------------------------------------------------
\subsection{Original Relational Specification Language}

For the specification of a relational property,
we initially proposed an extension~\cite{DBLP:conf/tacas/BlatterKGP17} of the \acsl specification
language with a new clause, \lstinline{relational}.
These clauses are attached to a function contract. A property
relating calls of different functions, such as \lstinline|R1| in
Figure \ref{fig:spec-ex1},
must appear in the contract of the last function involved
in the property, {\it i.e.} when all relevant functions are in scope.
In this new clause we introduced a new construct \lstinline{\call(f,<args>)},
denoting the value returned by the call \lstinline{f(<args>)}
to \lstinline{f} with arguments \lstinline{<args>}. This allows
relating several function calls in a \lstinline{relational} clause.
\lstinline{\call} can be used recursively, i.e. a parameter of a called
function can be the result of another function call.
In Figure \ref{fig:spec-ex1}, properties \lstinline{R1} and \lstinline{R2}
at lines 7--9 and 15--17 specify properties of
functions \lstinline{max} and \lstinline{min} respectively.

Note however that the \lstinline|\call| construct only allows speaking about
the return value of a C function. If the function has some side effects, there
is no way to express a relation between the values of memory locations that
are modified by distinct calls. Section~\ref{subsec:sideeffect} describes
the improvements that have been made to the initial version of the
relational specification language in order to support side effects.
To ensure that a function has no side effects, an \lstinline{assigns \nothing}
clause can be used.

%============================
\begin{figure}[tb]
  \lstset{basicstyle=\scriptsize\ttfamily,mathescape=true}
  %\begin{minipage}{0.5\columnwidth}
  \begin{subfigure}{0.47\textwidth}
\begin{lstlisting}[firstline=1, lastline=20]
/*@ requires x > INT_MIN;
    assigns  \nothing;
    behavior pos:
      assumes x >= 0;
      ensures \result  == x;
    behavior neg:
      assumes x < 0;
      ensures \result  == -x;*/
int abs (int x){
  return (x >= 0) ? x : (-x);
}

/*@ requires INT_MIN < x+y < INT_MAX;
    assigns \nothing;
    relational R1: \forall int x,y;
     \call(max,x,y) ==
      (x+y+\call(abs,x - y))/2; */
int max(int x,int y){
  return (x >= y) ? x : y;
}
\end{lstlisting}
    \vspace{-1mm}
    \caption{Original source code}
    \label{fig:spec-ex1}
  \end{subfigure}
  \hspace{2mm}
  %\end{minipage}
  %\begin{minipage}{0.5\columnwidth}
  \begin{subfigure}{0.49\textwidth}
\begin{lstlisting}[firstline=22, lastline=50]
/*@ axiomatic Relational_axiom {
  logic int max_acsl(int x, int y);
  logic int abs_acsl(int x);
  lemma Relational_lemma{L}:
    \forall int x, int y;
      max_acsl(x, y) ==
       ((x + y) + abs_acsl(x - y)) / 2;
  }*/

void relational_wrapper(int x, int y){
  int ret_var_1, ret_var_2;
  ret_var_1 = (x >= y) ? x : y;
  ret_var_2 = (x-y >= 0) ? x-y : (-(x-y));
  /*@ assert
       ret_var_1 ==
       ((x + y) + ret_var_2) / 2; */
  return;
}

/*@ assigns \nothing;
    behavior Relational_behavior:
      ensures
       \result ==
       max_acsl(\old(x), \old(y));
 */
int max(int x, int y){ ... }
\end{lstlisting}
    \vspace{-1mm}
    \caption{Excerpt of the code generated by \toolname}
    \label{fig:tran-ex1}
  \end{subfigure}
%  \vspace{-3mm}
  \lstset{basicstyle=\normalsize\ttfamily,}
  \caption{Pure function with relational properties}
%  \vspace{-3mm}
  \label{fig:ex1}
\end{figure}
%============================

%----------------------------------------------------------------------------
\subsection{Preprocessing of a Relational Property}

The previous work \cite{DBLP:conf/tacas/BlatterKGP17} also proposed a code transformation 
whose output can be
analyzed with standard deductive verification tools. This is materialized in
the \toolname plugin of \framac, that relies then on \Wp to prove the resulting
standard \acsl annotations.

Going back to our example, applying the transformation to property
\lstinline|R1| over function \lstinline|max|
gives the code of Figure~\ref{fig:tran-ex1}.
The generated code can be divided into three parts.
First, a new function, called \emph{wrapper}, is generated.
The wrapper function is inspired by the workaround proposed in \cite{BishopBC13}
and self-composition \cite{DBLP:journals/mscs/BartheDR11}.
As in self-composition, this wrapper function inlines the calls occurring in the
relational property under analysis, with a suitable renaming of local variables
to avoid interferences between the calls.

In addition, the wrapper records the results of the calls
in fresh local variables.
Then, in the spirit of calculational proofs~\cite{DBLP:conf/vstte/LeinoP13},
we state an assertion equivalent
to the relational property (lines 14--16 in Figure \ref{fig:tran-ex1}).
The proof of such an assertion is possible with a classic deductive
verification tool (\Wp with Alt-Ergo as back-end prover in our case).

However, the wrapper function only provides a solution
to prove relational properties. The ability to use these properties
as hypotheses in other proofs (relational or not)
must be reached otherwise. For this purpose, \toolname generates an ACSL 
axiomatic definition (cf. \lstinline{axiomatic} section at lines 1--8 in Figure \ref{fig:tran-ex1})
introducing a logical reformulation of the relational property as a lemma
(cf. lines 4--7) over otherwise unspecified logic functions
(\lstinline|max_acsl| and \lstinline|abs_acsl| in the example).
Furthermore, new postconditions are generated in the contracts of the C
functions involved in the relational property. They specify that there is
an exact correspondence between the original C function and its newly
generated logical \acsl counterpart.
Thanks to this axiomatic, POs over functions calling
\lstinline|max| and \lstinline|abs| will have the lemma in their environment
and thus will be able to take advantage of the proven relational property.
Note that the correspondence between
 \lstinline|max| and \lstinline|max_acsl|
(respectively \lstinline|abs| and \lstinline|abs_acsl|)
can only be done because \lstinline|max| and \lstinline|abs| do not access global memory
(neither for writing nor for reading). Indeed, since \lstinline|max_acsl| and \lstinline|abs_acsl| 
are pure logic functions, they do not have side effects and
their result only depends on their parameters.

%============================
\begin{figure}[tb]
  \lstset{basicstyle=\scriptsize\ttfamily,mathescape=true}
  %\begin{minipage}{0.5\columnwidth}
  \begin{subfigure}{0.49\textwidth}
\begin{lstlisting}[firstline=1, lastline=30]
/*@ assigns \nothing;*/
int Crypt(int m,int key){
  return m + key;
}

/*@ assigns \nothing;
    relational R3:
      \forall int m, key;
        \call(Decrypt,
          \call(Crypt,m,key),
          key)
        == m;*/
int Decrypt(int m,int key){
  return m - key;
}

/*@ assigns \nothing;
    ensures \result == m;
    relational R4:
     \forall int m,key;
      \call(run,
        \call(run,m,key),
        key)
      == m;*/
int run(int m,int key){
  int crypt, decrypt;
  crypt = Crypt(m,key);
  decrypt = Decrypt(crypt,key);
  return decrypt;
}
\end{lstlisting}
%    \vspace{-1mm}
    \caption{Original source code}
    \label{fig:spec-ex2}
  \end{subfigure}
  \hspace{2mm}
  %\end{minipage}
  %\begin{minipage}{0.5\columnwidth}
  \begin{subfigure}{0.49\textwidth}
\begin{lstlisting}[firstline=69, lastline=101]
/*@ axiomatic Relational_axiom {
  logic int run_acsl(int m, int key);

  lemma Relational_lemma{L}:
    \forall int m, int key;
    run_acsl(
      run_acsl(m, key),
      key)
    == m; }*/

void relational_wrapper(int m, int key){
  int tmp_1, tmp_2, tmp_3, tmp_4;
  tmp_1 = Crypt_aux_2(m,key);
  tmp_2 = Decrypt_aux_2(tmp_1,key);
  tmp_3 = Crypt_aux_2(tmp_2,key);
  tmp_4 = Decrypt_aux_2(tmp_3,key);
  /*@ assert tmp_4 == m;*/
  return; }

/*@ ensures \result == \old(m);
    assigns \nothing;
    behavior Relational_behavior:
      ensures \result ==
      run_acsl(\old(m), \old(key));*/
int run(int m, int key){
  int crypt;
  int decrypt;
  crypt = Crypt(m,key);
  decrypt = Decrypt(crypt,key);
  return decrypt; }
\end{lstlisting}
%    \vspace{-1mm}
    \caption{Transformed code}
    \label{fig:trans-ex2}
  \end{subfigure}
%  \vspace{-3mm}
  \lstset{basicstyle=\normalsize\ttfamily,}
  \caption{Functions \lstinline{Crypt} and \lstinline{Decrypt},
    used by function \lstinline{run}.}
%  \vspace{-3mm}
  \label{fig:ex2}
\end{figure}

To illustrate the use of relational properties in the proof of other
specifications, we can consider the postcondition and
property \lstinline{R4} of function \lstinline|run|
of Figure \ref{fig:spec-ex2} (inspired by the PISCO project\footnote{See \url{http://www.projet-pisco.fr/}.})
whose proof needs to use property \lstinline{R3}.
Thanks to their reformulation as lemmas and to the link between \acsl and C functions,
%(cf. lines 11--12, 18 of Figure \ref{fig:ex1}b for \lstinline{f1}),
\Wp automatically proves the assertion at line 17 (for property \lstinline{R4})
and the postcondition at line 20 of Figure \ref{fig:trans-ex2}.

\subsection{Soundness of the transformation}\label{sec:soundn-transf}

Since our transformation is introducing an ACSL \lstinline|axiomatic|, care must
be taken to avoid introducing inconsistencies in the specification. More
precisely, the \lstinline|axiomatic| specifies the intended behavior of the
ACSL counterpart of the C functions under analysis. The corresponding ACSL
functions are then only used in the contracts of those C functions.
In particular, since the wrapper is inlining the body of the functions
concerned by the relational property, the \lstinline|lemma| of the
\lstinline|axiomatic| cannot be used to prove the \lstinline|assert| annotation
inside the wrapper.

% Local Variables:
% compile-command: "pdflatex main.tex"
% mode: latex
% TeX-master: t
% TeX-PDF-mode: t
% mode: flyspell
% ispell-local-dictionary: "american"
% End:

%----------------------------
\section{Functions with Side Effects}
\label{subsec:sideeffect}
As mentioned above, the initial \toolname approach only works for
relational properties over pure functions.
More precisely, it allows proving relational properties of the form:
\begin{center}
\lstset{basicstyle=\scriptsize\ttfamily,}
\begin{scriptsize}
$\forall$ \ \lstinline{<args1>}, \dots, $\forall$ \lstinline{<argsN>}, \\
$P($ \lstinline'<args1>', \dots,\lstinline'<argsN>', \lstinline'\call(f_1,<args1>)', \dots, \lstinline{\call(f_N,<argsN>)}$)$
\end{scriptsize}
\lstset{basicstyle=\normalsize\ttfamily,}
\end{center}
for an arbitrary predicate $P$ invoking $N\ge 1$ calls of non-recursive
functions without side effects. In the context of the C programming language,
handling only pure functions is a major limitation.
We thus propose an extension of both
the specification language and the transformation technique in order to let
\toolname tackle a wider, more representative, class of C functions.

%----------------------------------------------------------------
\subsection{New Grammar for Relational Properties}

Relational properties are still introduced
by a \lstinline|relational| clause inside an \acsl contract. However, since
we might now refer to memory locations in either the pre- or the post-state of
any call implied in the relational property, we need to be able to make
explicit references to these states, and not only to the
value returned by a given call. Although more verbose,
the new syntax can also be used
for pure functions. For instance, property \lstinline|R1|
of Figure~\ref{fig:spec-ex1} can be rewritten as shown in
Figure~\ref{fig:new-grammar}.

%============================
\begin{figure}[tb]
  \lstset{basicstyle=\scriptsize\ttfamily,mathescape=true}
  \begin{lstlisting}[style=c,escapechar=§]
/*@ assigns \result \from x, y;
    relational R1:
       \forall int x1, y1;
          \callset(\call(max, x1, y1, id1),\call(abs, x1 - y1, id2)) ==>
           \callresult(id1) == (x1 + y1 + \callresult(id2)) / 2;
*/
int max(int x,int y) { ... }
  \end{lstlisting}
  \caption{Annotated C function with \lstinline{relational} annotations}
  \label{fig:new-grammar}
\end{figure}

%============================

More generally, we introduce the grammar shown in Figure \ref{fig:grammar}.
A relational clause is composed of three parts. First, we declare a set of
universally quantified variables, that will be used to express the
arguments of the calls that are related by the clause.
Then, we specify the set of calls on which we will work in
the \emph{relational-def} part. As shown in Figure~\ref{fig:grmmapred},
each call is then associated to an identifier \emph{call-id}. 
In the property \lstinline'R1' of Figure~\ref{fig:new-grammar},
two function calls are explicitly specified in the
\lstinline'\callset' construct and not directly in the predicate.
Each call has its own identifier (\lstinline{id1} and \lstinline{id2}
respectively). Finally, the relational property itself is given as an
\acsl predicate in the
\emph{relational-pred} part. As described in Figure~\ref{fig:grmmapred},
in addition to standard \acsl constructs, three new terms can be used. First,
\lstinline|\callpure| can be used to indicate the value returned by a pure
function as was done with the \lstinline|\call| built-in in the original
version of \toolname. This allows specifying relational properties over pure
functions without the overhead required for handling side-effects. As before,
nested \lstinline|\callpure| are allowed.
Second, \lstinline|\callresult|, as used in Figure~\ref{fig:new-grammar},
takes a \emph{call-id} as parameter and refers
to the value returned by the corresponding call in \emph{relational-def}.
Finally, each such \emph{call-id} gives rise to two logic labels.
Namely, \lstinline|Pre_|\emph{call-id} refers to the
pre-state of the corresponding
call, and \lstinline|Post_|\emph{call-id} to its post-state. These labels can
in particular be used in the \acsl term
\lstinline|\at(e,$L$)| that indicates that the
term \lstinline|e| must be evaluated in the context of the program
state linked to logic label $L$.
Figure~\ref{fig:side-spec} below shows an example of their use.

\setlength{\grammarparsep}{8pt plus 1pt minus 1pt}
\setlength{\grammarindent}{6em}

\begin{figure}[tb]
\begin{scriptsize}
\begin{subfigure}{0.49\textwidth}
\setlength{\grammarindent}{1.5em}
  \begin{grammar}

<call-id> ::= id

<bin-rel> ::= "==" | "!=" | "<=" | ">=" | ">" | "<"

<function-parameter> ::= <relational-call-terms>+

<function-name> ::= poly-id

<function-call> ::=
  "\\call("<inlining-option>","\\
   <function-name>","\\
   <function-parameter>","\\
   <call-id>")"

<call-parameter> ::=  <function-call>+

<relational-def> ::=  "\\callset("<call-parameter> ")"

<relational-pred> ::= "\\true" | "\\false"
\alt <relational-terms> <bin-rel> <relational-terms>
\alt <relational-pred> "&&" <relational-pred>
\alt <relational-pred> "||" <relational-pred>
\alt <relational-pred> "==>" <relational-pred>
\alt "!"<relational-pred>
\alt "\\forall" <binders> ";"<relational-pred>
\alt "\\exists" <binders> ";"<relational-pred>

<relational-annot> ::= "relational" <relational-clause>

<relational-clause> ::= "" \\
"\\forall" <binders> ";" "" \\
<relational-def> "==>" <relational-pred>

  \end{grammar}
%  \vspace{-4mm}
  \caption{Grammar of relational predicates}
  \label{fig:grmmapred}
\end{subfigure}
\begin{subfigure}{0.49\textwidth}
\setlength{\grammarindent}{1.5em}
\begin{grammar}

<literal> ::= "\\true" | "\\false" | int | float

<relational-label> ::= "Post_"<call-id>
\alt "Pre_"<call-id>

<bin-op> ::= "+" |  "-" | "*" | "/" |

<result-reference> ::= "\\callresult(" <call-id> ")"

<pure-function-parameter> ::= <relational-call-terms>+

<inlining-option> ::= int

<pure-function-name> ::= poly-id

<pure-function-call> ::= "" \\
"\\callpure("<inlining-option>"," <pure-function-name>","<pure-function-parameter>")"

<relational-call-terms> ::= <literal>
\alt <pure-function-call>
\alt <relational-call-terms> <bin-op> <relational-call-terms>

<relational-terms> ::= <literal>
\alt <relational-terms> <bin-op> <relational-terms>
\alt <result-reference>
\alt "\\ at("<poly_id> "," <relational-label> ")"
\alt <pure-function-call>

  \end{grammar}
%\vspace{-4mm}
\caption{Grammar of relational terms}
\label{fig:grmmaterm}
\end{subfigure}
\end{scriptsize}
\caption{Grammar for relational properties}
%\vspace{-0.7cm}

\label{fig:grammar}
\end{figure}

%-----------------------------------------------------------------------------
\subsection{Global Variables Accesses}

As said before, the new syntax for relational properties enables us to
speak about the value of global variables at various states of the execution,
thanks to the newly defined logic labels bound to each call involved in the
\lstinline|\callset| of the property. This is for instance the case in
the relational property of Figure~\ref{fig:side-spec}, which indicates that
\lstinline|h| is monotonic with respect to \lstinline|y|, in the sense
that if a first call to \lstinline|h| is done in a state \lstinline|Pre_id1|
where the value of \lstinline|y| is strictly less than in the pre-state
\lstinline|Pre_id2| of a second call, this will also be the case in the
respective post-states \lstinline|Post_id1| and \lstinline|Post_id2|.

%============================
\begin{figure}[tb]
  \lstset{basicstyle=\scriptsize\ttfamily,mathescape=true}
  \begin{subfigure}{0.51\textwidth}
    \begin{lstlisting}[style=c,escapechar=§]
int y;

/*@ assigns y \from y;
    relational R1:
     \callset(\call(h,id1),
              \call(h,id2))
     ==>
     \at(y,Pre_id1) < \at(y,Pre_id2)
     ==>
     \at(y,Post_id1) < \at(y,Post_id2);
*/
void h(){
  int a = 10;
  y = y + a;
  return;
}
    \end{lstlisting}
%    \vspace{-4mm}
    \caption{Annotated C function with \lstinline{relational} \\ annotations}
    \label{fig:side-spec}
  \end{subfigure}
%  \hspace{2mm}
  \begin{subfigure}{0.48\textwidth}
    \begin{lstlisting}[style=c,escapechar=§]
int y;

/*@ axiomatic Relational_axiom_1 {
  predicate 
     h_acsl(int y_pre, int y_post);

  lemma Relational_lemma_1:
  \forall int y_id2_pre, y_id2_post,
              y_id1_pre, y_id1_post;
   h_acsl(y_id2_pre, y_id2_post)
   ==> h_acsl(y_id1_pre, y_id1_post)
   ==> y_id1_pre < y_id2_pre
   ==> y_id1_post < y_id2_post; }*/

/*@ assigns y \from y;
    behavior Relational_behavior_1:
      ensures h_acsl(\at(y,Pre), 
      \at(y,Post));*/
void h(void){ ... }

int y_id1;
int y_id2;

void relational_wrapper_1(void){
    int a_1 = 10;
    y_id1 = y_id1 + a_1;
    int a_2 = 10;
    y_id2 += y_id2 + a_2;
  /*@ assert Rpp:
    \at(y_id1,Pre) < \at(y_id2,Pre) ==>
    \at(y_id1,Here) < \at(y_id2,Here);*/
  return;
}
    \end{lstlisting}
    \caption{Transformed code for verification and use of relational properties with side effect}
    \label{fig:side-transformation}
  \end{subfigure}
%  \vspace{-3mm}
  \caption{Relational property on a function with side-effect}
%  \vspace{-0.7cm}
\end{figure}
%============================

Generation of the wrapper function is more complicated in presence of
side-effects. As presented in~\cite{DBLP:journals/mscs/BartheDR11},
each function call must operate on its own memory state, separated
from the other calls in order for self-composition to work.
We thus create as many duplicates of global variables as
needed to let each part of the wrapper use its own set of copies.
However, to avoid useless copies, \toolname requires that each
function involved in a relational property has been equipped with a proper
set of ACSL \lstinline|assigns| clauses,
including \lstinline|\from| components. This constraint is similar
to what is proposed in~\cite{DBLP:journals/sttt/CuoqMPP11}, and ensures that
only the parts of the global state that are accessed (either for writing or
for reading) by the functions under analysis are subject to duplication.
As an example, the wrapper function corresponding to our \lstinline|h|
function of Figure~\ref{fig:side-spec} is shown in lines 24--33 of
Figure~\ref{fig:side-transformation}.

Finally, the generated axiomatic definition
enabling the use of the relational property in other POs must also be modified.
The original transformation uses a logic function that is supposed to
return the same \lstinline|\result| as the C function. However,
since logic functions are always pure,
this mechanism is not sufficient to characterize side effects
in the logic world.
Instead, we declare a predicate that takes as parameters not only the
returned value and the formal parameters of the C function, but also the relevant
parts of the program states that
are involved in the property. As for the wrapper function, these additional
parameters are inferred from the \lstinline|assigns ... \from ... | clauses
of the corresponding C functions. For instance, predicate \lstinline|h_acsl|,
on line 5 of figure~\ref{fig:side-transformation}, takes two arguments
representing the values of \lstinline|y| before and after and execution of
\lstinline|h|. This link between the \acsl predicate and the C function is
again materialized by an \lstinline|ensures| clause (lines 17--18).
The lemma defining the \acsl predicate is more
complex too, since we have to quantify over the values of all the global
variables at all relevant program states. In the example, this is shown on
lines 7--13, where we have 4 quantified variables representing the value
of global variable \lstinline|y| before and after both calls involved in the
relational property.

%%--------------------------------------------------------------------------------------------------
\subsection{Support of Pointers}

In the previous section, we have shown how to specify relational
properties in presence of side effects over global variables,
and how the transformations for both proving and using
a property are performed. However, support of pointer dereference
is more complicated. Again, as proven in~\cite{DBLP:journals/mscs/BartheDR11}
Self-Composition works if the memory footprint of each
call is separated from the others. Thus,
in order to adapt our method, we must ensure that
pointers that are accessed during two distinct calls point to different
memory locations. As above, such accesses are given by
\lstinline|assigns ... \from ...| clauses in the contract of the corresponding
C functions. An example of a relational property on a function \lstinline|k|
using pointers (monotonicity with respect to the content of a pointer)
is given in Figure~\ref{fig:pointer-example}, where \lstinline|k| is
specified to assign \lstinline|*y| using only its initial content.

Memory separation is enforced using \acsl's built-in predicate
\lstinline|\separated|. For the wrapper function, we add a
\lstinline|requires| clause stating the appropriate \lstinline|\separated|
locations. This can be seen on
Figure~\ref{fig:pointer-transformation}, line 20, where we request that the
copies of pointer \lstinline|y| used for the inlining of both calls to
\lstinline|k| points to two separated area in the memory.
Similarly, in the axiomatic part, the lemma adds separation constraints over
the universally quantified pointers (line 9 in the Figure~\ref{fig:pointer-transformation}).

We also need to refine the declaration of the
predicate in presence of pointer accesses.
First, the predicate now needs to explicitly take as parameters the pre- and
post-states of the C function. In \acsl, this is done by specifying
\emph{logic labels} as special parameters, surrounded by braces, as shown
in line 3 of Figure~\ref{fig:pointer-transformation}. Second, a \lstinline'reads' clause allows one to specify
the footprint of the predicate, that is, the set of memory accesses that
the validity of the predicate depends on (line 4). Similarly, the lemma
on lines 6--13 takes 4 logic labels as parameters, since it relates two calls
to \lstinline|k|, each of them having a pre- and a post-state.

%============================
\begin{figure}[tb]
  \vspace{-0.5cm}
  \lstset{basicstyle=\scriptsize\ttfamily,mathescape=true}
  \begin{subfigure}{0.36\textwidth}
    \begin{lstlisting}[style=c,escapechar=§]
/*@ assigns *y \from *y;
    relational R1:
    \callset(
      \call(k,id1),
      \call(k,id2))
     ==>
       \at(*y,Pre_id1) <
       \at(*y,Pre_id2)
       ==>
        \at(*y,Post_id1) <
        \at(*y,Post_id2);
*/
void k(int *y){
  *y = *y + 1;
  return;
}
    \end{lstlisting}
%    \vspace{-4mm}
    \caption{Original annotated C function}
    \label{fig:pointer-example}
  \end{subfigure}
  \hspace{4mm}
  \begin{subfigure}{0.60\textwidth}
    \begin{lstlisting}[style=c,escapechar=§]
/*@ axiomatic Relational_axiom_1 {
predicate 
k_acsl{pre, post}(int *y)
  reads \at(*y,post), \at(*y,pre);

lemma Relational_lemma_1
  {pre_id2, post_id2, pre_id1, post_id1}:
\forall int *y_id2, int *y_id1;
\separated(y_id1,y_id2)
  ==> k_acsl{pre_id2, post_id2}(y_id2)
  ==> k_acsl{pre_id1, post_id1}(y_id1)
  ==> \at(*y_id1,pre_id1) < \at(*y_id2,pre_id2)
  ==> \at(*y_id1,post_id1) < \at(*y_id2,post_id2);
}*/
/*@ assigns *y \from *y;
    behavior Relational_behavior_1:
      ensures k_acsl{Pre, Post}(y);*/
void k(int *y){  ... }

/*@ requires \separated(y_id1, y_id2);*/
void relational_wrapper_1(int *y_id1, int *y_id1){
  *y_id1 = *y_id1 + 1;
  
  *y_id2 = *y_id2 + 1;
  
  /*@ assert Rpp:
  \at(*y_id1,Pre) < \at(*y_id2,Pre) ==>
  \at(*y_id1,Here) < \at(*y_id2,Here);*/
  return;
}
    \end{lstlisting}
%    \vspace{-4mm}
    \caption{Code transformation}
    \label{fig:pointer-transformation}
  \end{subfigure}
%  \vspace{-3mm}
  \caption{Relational property in presence of pointers}
%  \vspace{-0.7cm}
\end{figure}
%============================
It should be noted that the memory separation assumption makes the tool 
verify relational properties without pointer aliasing.
Support of properties with pointer aliasing is left as future work.

%%--------------------------------------------------------------------------------------------------

% Local Variables:
% compile-command: "pdflatex main.tex"
% mode: latex
% TeX-master: t
% TeX-PDF-mode: t
% mode: flyspell
% ispell-local-dictionary: "american"
% End:

%----------------------------
\section{Recursive Functions}
\label{subsec:recursive}
We have shown in the previous section how we handle functions with 
side effects. Let us now focus on another class of functions,
namely recursive functions. Support for recursive functions in \toolname
is interesting because it is very natural to specify such functions with
relational properties. For example, a naive
specification of a \lstinline{fact} function computing
the factorial of an integer can be written as
\begin{equation*}
  \begin{cases}
    \forall x.\ x \leq 1 \implies fact(x) = 1,\\
    \forall x.\ x > 1 \implies fact(x) = fact(x-1) * (x)
  \end{cases}
\end{equation*}

The corresponding relational properties are given in Figure~\ref{fig:rec-spec}.
The proof of the \lstinline|Induction| property requires a modification to
the generation of the wrapper function, that can be observed in
Figure~\ref{fig:rec-transformation}. Indeed, we do not want to inline the
second call to \lstinline|fact| on line~12, in order to take advantage of
the fact that, since \lstinline|fact| is a pure function that does not read
anything from the global environment, this call returns the same value as
the one of line~9, obtained by inlining the call to \lstinline|fact(x1)|.
This is why, as was indicated on Figure~\ref{fig:grammar}, there is an
optional argument to the \lstinline|\callpure| construct,
that indicates the maximal depth that the inlining can reach in the wrapper.
The default value of \lstinline|1|, which is also used explicitly
in our example for the first call, on line 9 of Figure~\ref{fig:rec-spec},
means that we inline the body of the function
once (i.e. if the function calls other functions, including itself, these
calls themselves will not be inlined). When this parameter is set to
\lstinline|0|, as is the case for the second call in our example (line 10),
we keep the call as such in the wrapper.

%============================
\begin{figure}[tb]
%  \vspace{-0.7cm}
  \lstset{basicstyle=\scriptsize\ttfamily,mathescape=true}
  \begin{subfigure}{0.46\textwidth}
    \begin{lstlisting}[style=c,escapechar=§]
/*@ assigns \result \from x;
    relational Base:
      \forall int x1;
        x1 <= 1 ==>
          \callpure(1,fact,x1) == 1;
    relational Induction:
      \forall int x1;
        x1 > 1 ==>
        \callpure(1,fact,x1) ==
          \callpure(0,fact,x1-1)*x1;
*/
int fact(int x) {
	if(x <= 1){
		return 1;
	}
	else{
		return fact(x-1)*x;
	}
}
    \end{lstlisting}
%    \vspace{-4mm}
    \caption{Annotated recursive C function with \lstinline{relational} clauses}
    \label{fig:rec-spec}
  \end{subfigure}
  \hspace{5mm}
  \begin{subfigure}{0.46\textwidth}
    \begin{lstlisting}[style=c,escapechar=§]
void relational_wrapper_2(int x1){
int return_var_rela_2;
int return_var_rela_3;
{
  if (x1 <= 1) {
    return_var_rela_2 = 1;
  }
  else {
    return_var_rela_2 = fact(x1-1)*x1;
  }
}
return_var_rela_3 = fact(x1-1);
/*@ assert Rpp:
     x1 > 1 ==>
     return_var_rela_2 ==
     return_var_rela_3*x1;
*/
return;
}
    \end{lstlisting}
    \caption{Code transformation for the proof of the second relational property}
    \label{fig:rec-transformation}
  \end{subfigure}
%  \vspace{-3mm}
  \caption{Relational property on recursive C function without side effects}
%  \vspace{-0.5cm}
\end{figure}

Support for recursive functions is not limited to pure functions.
Recursive functions with side effects can also be handled.
In particular, as shown in the grammar,
each \lstinline|\call| appearing in a \lstinline|\callset| can also have
an inlining directive.
For instance,
we can consider another implementation of
the factorial, whose result is this time recorded in a global variable
\lstinline'r' (Figure~\ref{fig:rec-side-effect}).
The corresponding relational properties (lines 5--9) are
similar to the pure case. However, the proof is slightly different, since
the function has side effects, we cannot use logic function
equality. Instead, we use the relational property
as an induction hypothesis and inline both functions.

\begin{figure}[tb]
  \lstset{basicstyle=\scriptsize\ttfamily,mathescape=true}
  \begin{lstlisting}[style=c,escapechar=§]
int r;

/*@ requires x >= 0;
    assigns r \from r,x;
    relational \forall int x1;
     \callset(\call(1,fact,x1,id1)) ==> x1 <= 1 ==> \at(r,Post_id1) == 1;
    relational \forall int x1;
     \callset(\call(1,fact,x1,id2), \call(1,fact,x1-1,id3))
     ==> x1 > 1 ==> \at(r,Post_id2) == \at(r,Post_id3)*x1;
*/
void fact(int x) {
  if(x <= 1){
    r = 1;
    return;
  }
  else{
    fact(x-1);
    r = r * x;
    return;
 }
}
  \end{lstlisting}
%  \vspace{-4mm}
  \caption{Relational property on recursive C function with side effects}
  \label{fig:rec-side-effect}
\end{figure}

Note that in this case, a call to the function itself appears in the wrapper,
contrarily to the situation detailed in section~\ref{sec:soundn-transf}.
However, under the assumption that the function always terminates, this call is
performed on arguments that are strictly smaller than the ones of the wrapper
itself. Hence, the \lstinline|axiomatic| can be used as an induction hypothesis
in the sense that the wrapper allows us to prove that if the relational
property holds for arguments smaller than \lstinline|x|, then it holds for
\lstinline|x|.

% Local Variables:
% compile-command: "pdflatex main.tex"
% mode: latex
% TeX-master: t
% TeX-PDF-mode: t
% mode: flyspell
% ispell-local-dictionary: "american"
% End:

%----------------------------
\section{llustrative Examples}
\label{subsec:example}
We have seen how to express
relational properties over a large class of C functions and how \toolname
can generate C code and plain \acsl specifications for proving and using these
properties through a standard \Wp process.
To check that this approach works in practice, we have tested our tool
on different benchmarks. These tests aim at confirming:
\begin{itemize}
\item the ability to specify various relational properties over a large
class of functions;
\item the capacity to prove and use such properties using the
  generated transformation;
\item the support of a large range of function implementations;
\item the ability to use other techniques (runtime checks, test generation
  for invalidating the property) when \Wp fails to discharge a corresponding
  PO.
\end{itemize}
The first subsection will present our own benchmark composed of a mix of
different types of relational properties. This benchmark is mainly designed to
validate the two first items. The second subsection will show how \toolname
has performed on the benchmark proposed in \cite{SousaD16}.
This will confirm the second and third points. Finally, we will present
in Section~\ref{subsec:test} our use of the \eacsl and \stady plugins
assessing the last point.

%----------------------------------------------------------------------------------------------------
\subsection{Internal Examples}
As stated previously, we have tested \toolname on a set of
relational properties extracted from real case studies.
This includes in particular encryption, as presented in
Section~\ref{subsec:rpp}, monotonicity (Section~\ref{subsec:sideeffect})
or the factorial of Section~\ref{subsec:recursive}, but also properties
found in map/reduce, as the one in row~\ref{ex:map-reduce} in Figure\ref{tab:rpp-bench},
stating that the choice of the partitioning for the initial set of data
should not play a role in the final result.
The benchmark is also composed of more academic examples like
linear algebraic properties of matrices, over functions containing
loops (rows~\ref{ex:sum-trans} and~\ref{ex:det-trans}), or
the property of row~\ref{ex:med}, that states the symmetry of the median of
three numbers.

Figure~\ref{tab:rpp-bench} summarizes the results obtained on the benchmark.
The first three columns indicate respectively whether the
corresponding property could be specified and the corresponding code
transformation generated, proved and used as an hypothesis in
other proofs. The last three columns show what kind of C constructs are used
in the implementation of the functions under analysis, namely side effects,
presence of loops (which are always difficult for \Wp-related
verification techniques, due to the need for loop invariants),
and presence of recursive functions.

\newcounter{exampleno}
\newcommand{\exampleref}[1]{
  \refstepcounter{exampleno}\theexampleno\label{ex:#1}
}
\begin{figure}[tb]
\renewcommand{\tabcolsep}{0.1cm}
\renewcommand{\arraystretch}{0.4}
\begin{scriptsize}
  \begin{tabular}{ c  c | c c c | c c c }
    Num & Relational Property &
    \begin{tabular}{@{}c@{}}
      \\ Specified / \\ Generated \\ ~%
     \end{tabular}
    & Verified & Used &
    \begin{tabular}{@{}c@{}}
      \\ Side \\ effect \\ ~%
    \end{tabular}
    & Loop & Recursive\\
    \hline
    \exampleref{monotone} &
    \begin{tabular}{@{}c@{}}
      \\ $\forall x1,x2 \in \mathbb{Z}:$ \\ $\ x1 < x2 \Rightarrow f(x1) < f(x2)$ \\ ~%
    \end{tabular}
    &
    \cmark & \cmark & \cmark & \cmark & \xmark & \xmark\\
    \exampleref{fact} &
    \begin{tabular}{@{}c@{}}
      \\  $ \forall \ x;$ \\ $f(x+1) = f(x)*(x+1)$ \\~%
    \end{tabular}
    &
    \cmark & \cmark & \cmark & \cmark & \xmark & \cmark\\
    \exampleref{order} &
    \begin{tabular}{@{}c@{}}
      \\ $\forall x, \  f_1(x) \leq f_2(x) \leq f_3(x)$\\~%
    \end{tabular}
    &
    \cmark & \cmark & \xmark & \xmark & \xmark & \xmark\\
    \exampleref{idempotent} &
    \begin{tabular}{@{}c@{}}
      \\ $\forall x, f(f(x)) = f(x)$ \\~%
    \end{tabular}
    &
    \cmark & \cmark & \xmark & \xmark & \cmark & \xmark\\
    \exampleref{decrypt} &
   \begin{tabular}{@{}c@{}}
      \\ $\forall \ Msg,Key;$ \\ $Decrypt(Encrypt(Msg,Key),Key)=Msg$ \\ ~%
    \end{tabular}
    &
    \cmark & \cmark & \cmark & \cmark & \cmark & \xmark\\
    \exampleref{map-reduce} &
    \begin{tabular}{@{}c@{}}
      \\  $ \forall \ t,sub_{t1}, ..., sub_{tn};$
      \\$t = sub_{t1} \cup ... \cup sub_{tn} \Rightarrow$ \\
      $ max(t) = max(max(sub_{t1}),...,max(sub_{tn}))$ \\~%
    \end{tabular}
    &
    \cmark & \cmark & \xmark & \cmark & \cmark & \xmark\\
    \exampleref{sum-trans} &
    \begin{tabular}{@{}c@{}}
      \\  $ \forall \ A,B;$ \\ $(A + B)^T = (A^T + B^T)$ \\~%
    \end{tabular}
    &
    \cmark & \cmark & \xmark & \xmark & \cmark & \xmark\\
    \exampleref{det-trans} &
    \begin{tabular}{@{}c@{}}
      \\  $\mathrm{det}(A) = \mathrm{det}(A^\intercal)$ \\~%
    \end{tabular}
    &
    \cmark & \cmark & \xmark & \xmark & \cmark & \xmark\\
    \exampleref{left-lin} &
    \begin{tabular}{@{}c@{}}
      \\  $ \forall x1,x2,y ,f(x1+x2,y) = f(x1,y) + f(x2,y)$ \\~%
    \end{tabular}
    &
    \cmark & \cmark & \cmark & \xmark & \xmark & \cmark\\
    \exampleref{med} &
    \begin{tabular}{@{}c@{}}
      \\  $ \forall a,b,c, \ \mathrm{Med}(a,b,c) = \mathrm{Med}(a,c,b)$ \\~%
    \end{tabular}
    &
    \cmark & \cmark & \xmark & \xmark & \xmark & \xmark\\
  \end{tabular}
\end{scriptsize}
\caption{Summary of relational properties considered by \toolname}
\label{tab:rpp-bench}
\end{figure}
%\vspace*{\fill}
%--------------------------------------------------------------------
\subsection{Comparator Functions}
We also evaluated \toolname on the benchmark proposed
in \cite{SousaD16}. It is composed of a collection of
flawed and corrected implementations of comparators over a variety of
data types written in Java, inspired from a collection of
Stackoverflow~\footnote{\url{https://stackoverflow.com}} questions.
Translating the Java code into C was straightforward and fully
preserved the semantics of the functions.
We focused on the same properties as \cite{SousaD16}, that is
anti-symmetry (P1), transitivity (P2) and extensionality (P3). Mathematically,
these properties can be expressed as such:

\begin{gather*}
P1: \forall\ s1,s2.\ compare(s1,s2) = -compare(s2,s1)\\
P2: \forall\ s1,s2,s3.\ compare(s1,s2) > 0 \land compare(s2,s3) > 0 \\
\Rightarrow compare(s1,s3) > 0\\
P3: \forall\ s1,s2,s3.\ compare(s1,s2) = 0 \Rightarrow
(compare(s1,s3) = compare(s2,s3))\\
\end{gather*}

Results are depicted in Table~\ref{tab:compare-bench}. For each comparator,
we indicate whether the properties P1, P2 and P3 hold according to \toolname
(\cmark and \xmark show whether the property was proved valid
by \Wp).
We get similar results as \cite{SousaD16},
with the exception of PokerHand, for which the generated wrapper function
seems currently out of reach for \Wp (limits of scalability due to the combinatorial explosion 
of self-composition). However, by rewriting the function
in a more modular way,
\Wp was able to handle the example.

\begin{figure}[!htb]
\renewcommand{\tabcolsep}{0.5cm}
\begin{tabular}{l | c c c | c c c}
 & \multicolumn{3}{c|}{Proof (\Wp)} & \multicolumn{3}{c}{Counterex. gen. (\stady)}\\
Benchmark & P1 & P2 & P3 & P1 & P2 & P3\\
\hline
ArrayInt-false.c & \cmark & \cmark & \xmark & \cko & \cko & \cmark\\
ArrayInt-true.c & \cmark & \cmark & \cmark & \cko & \cko & \cko \\
CatBPos-false.c & \xmark & \xmark & \xmark & \cmark & \cmark & \cmark \\
Chromosome-false.c & \cmark & \xmark & \xmark & \cko & \ctimeout & \cmark \\
Chromosome-true.c & \cmark & \cmark & \cmark & \cko & \cko  & \cko\\
ColItem-false.c & \xmark & \xmark & \xmark & \cmark & \cmark & \cmark\\
ColItem-true.c & \cmark & \cmark & \cmark & \cko & \cko & \cko\\
Contact-false.c & \cmark & \xmark & \xmark & \cko & \cmark & \cmark\\
Container-false-v1.c & \xmark & \cmark & \cmark & \cmark & \cko & \cko\\
Container-false-v2.c & \xmark & \xmark & \xmark & \cmark & \cmark &\cmark\\
Container-true.c & \cmark & \cmark & \cmark & \cko & \cko & \cko\\
DataPoint-false.c & \xmark & \xmark & \xmark & \cmark & \cmark & \cmark\\
FileItem-false.c & \cmark & \cmark & \xmark & \cko & \cko  & \cmark\\
FileItem-true.c & \cmark & \cmark & \cmark & \cko & \cko & \cko\\
IsoSprite-false-v1.c & \xmark & \xmark & \xmark & \cmark & \cmark & \cmark\\
IsoSprite-false-v2.c & \xmark & \xmark & \cmark & \cmark & \cmark & \cko\\
Match-false.c & \xmark & \cmark & \xmark & \cmark & \cko & \cmark\\
Match-true.c & \cmark & \cmark & \cmark & \cko & \cko & \cko\\
NameComparator-false.c & \xmark & \cmark & \cmark & \cmark & \cko & \cko\\
NameComparator-true.c & \cmark & \cmark & \cmark & \cko & \cko & \cko\\
Node-false.c & \cmark & \cmark & \xmark & \cko & \cko & \cmark\\
Node-true.c & \cmark & \cmark & \cmark & \cko & \cko & \cko\\
NzbFile-false.c & \xmark & \cmark & \cmark & \cmark & \cko & \cko\\
NzbFile-true.c & \cmark & \cmark & \cmark & \cko & \cko & \cko\\
PokerHand-false.c & \cmark & \xmark & \xmark & \cko & \ctodo & \ctodo\\
PokerHand-true.c & \cmark & \cmark & \cmark & \cko & \cko & \cko\\
Solution-false.c & \cmark & \cmark & \xmark & \cko & \cko & \cmark\\
Solution-true.c & \cmark & \cmark & \cmark & \cko & \cko & \cko\\
TextPosition-false.c & \cmark & \xmark & \xmark & \cko & \cmark & \cmark\\
TextPosition-true.c & \cmark & \cmark & \cmark & \cko & \cko & \cko\\
Time-false.c & \xmark & \cmark & \cmark & \cmark & \cko & \cko\\
Time-true.c & \cmark & \cmark & \cmark & \cko & \cko & \cko\\
Word-false.c & \xmark & \xmark & \cmark & \cmark & \cmark & \cko\\
Word-true.c & \cmark & \cmark & \cmark & \cko & \cko & \cko\\
\end{tabular}
\caption{Comparator properties analysed with \Wp and \stady after \toolname translation}
\label{tab:compare-bench}
%\vspace{-1cm}
\end{figure}

%----------------------------
\section{Dynamic Verification}
\label{subsec:test}
\subsection{Counterexample Generation}
For the properties that do not hold in the comparator benchmark,
we have been able to find
counterexamples thanks to the proposed encoding of a relational property
by self-composed code and using another
\framac plugin, \stady~\cite{DBLP:conf/tap/PetiotKBGJ16}.
\stady\footnote{See \url{https://github.com/gpetiot/Frama-C-StaDy}}
is a testing-based counterexample generator. In particular, \stady tries to
find an input vector that will falsify an \acsl annotation for which \Wp could
not decide whether it holds, thereby showing that the code is not conforming
to the specification.

We apply \stady to try to find a test input %for an execution path
such that the
\lstinline|assert| clause at the end of the \lstinline|wrapper| function is false.
The results are shown in the \stady columns of Figure~\ref{tab:compare-bench}.
Obviously, \stady does not try to find counterexamples
for properties that are proved valid by \Wp. For properties that
are not proved valid, \cmark indicates that a counterexample is
found (within a timeout of 30 seconds),
while \ctimeout indicated the only case where a counterexample is not
generated before a 30-second timeout.
A longer timeout (60 minutes) did not improve the situation in that case.
Symbol \ctodo denotes two cases where the code translation
uses features that are currently not yet supported by \stady.
As shown in the table, thanks to the \toolname translation,
\stady was able to find counterexamples for
almost all unproven properties. Notice that some examples required
minor modifications so that \stady can be used.
To be able to use testing, we had of course to add bodies for unimplemented functions.
Other modifications consisted %for example
in reducing the input space to a representative smaller domain
(by limiting the size of an input array)
for some examples to facilitate counterexample generation~\cite{DBLP:conf/tap/PetiotKBGJ16}.
%or changing datatypes. %NK pourquoi ?

\subsection{Runtime Assertion Checking}
The code transformation technique of \toolname also enables runtime verification
of relational properties through the
\eacsl plugin~\cite{Delahaye/SAC13,VorobyovKS-e-acsl-segmodel}.
More precisely,
the \eacsl plugin translates \acsl annotations into C code that will check them
at runtime and abort
execution if one of the annotations fails.
We tested the \eacsl plugin on 
%, in particular,
%to generate executables by calling the
%\lstinline|wrapper| function
%generated by \toolname in a main function
the test inputs generated by \stady in order to check that each generated counterexample does
indeed violate the relational property.
As expected, the obtained results validate those of the previous section.
Since counterexample generation with \stady~\cite{DBLP:conf/tap/PetiotKBGJ16}
basically includes a runtime assertion checking step for each test datum
considered during the test generation process, we do not present
the results of this step in separate columns.

%-------------------------------------------------------------
\section{Conclusion and Future Work}
\label{sec:conclusion}
We have presented a major extension
to an existing verification technique for relational properties,
implemented in the \framac plugin \toolname. The extension adds support for
functions with side effects (access to global variables and pointer dereferences)
and recursive functions. \toolname relies on  \framac/\Wp
for automatic or interactive proof of the relational properties
and offers the ability to use them as hypothesis in other proofs.
Moreover, beyond \Wp, \toolname also allows users to take advantage of \eacsl and \stady 
plugins to verify relational properties at runtime
and to produce a test input exhibiting the issue when a function does not
respect the specified relational property.
We have also shown that our implementation can handle a wide variety of
properties and code:
we consider a large class of relational properties
with several, possibly nested, function calls.

However, there are still some limitations, inherent to
our use of sequential self-composition.
First, in the case of relational properties linking functions with
large bodies or a large number of functions, the size of the generated
wrapper function may explode, leading to POs that cannot be handled by automated
theorem provers or even generated by weakest precondition calculus.
A first solution for this problem is to use the modularity
of the approach to reduce the size of the function and prove sub-properties.
However, it is not always possible to modify an existing implementation.
Alternative methods, based on a generalization of the technique proposed
in~\cite{DBLP:journals/sttt/CuoqMPP11} for verifying \lstinline|\from| clauses,
and that do not rely on the generation of a wrapper function seem thus
desirable.
The notation of relational properties in the presence of side effects
can be seen somewhat heavy to use.
To make this notation more succinct, 
some shorthands for most common usages will be useful.
The possibility to use runtime verification and testing is an important benefit in situations
where the proof does not conclude.
Furthermore, treatment of loops needs to be improved. In particular,
it is not possible yet to specify ``relational invariants'' that would allow
relating the behavior of a loop in two different contexts, while this is
often necessary to complete the proof of a relational property. Solutions based
on program products~\cite{prograprodu} look promising. Finally, as already
mentioned, we need to extend our technique to handle potential aliases across
the executions involved in a relational property.

%as well as applying \toolname for verification of IoT software (e.g. for wireless sensor networks).

% Local Variables:
% compile-command: "pdflatex main.tex"
% mode: latex
% TeX-master: t
% TeX-PDF-mode: t
% mode: flyspell
% ispell-local-dictionary: "american"
% End:

%\begin{small}
\medskip
\noindent
%\paragraph{Acknowledgment.}
\textit{Acknowledgment.} %TODO
%% Part of the research work leading to these results has received funding
%% for DEWI project (www.dewi-project.eu) from the ARTEMIS
%% Joint Under\-ta\-king under grant agreement No. 621353, 
%% and for the S3P project from French DGE and BPIFrance.
The authors thank the \framac and \pathcrawler teams for providing the
tools and support.
Special thanks to Fran\c{c}ois Bobot, Lo\"{i}c Correnson,
and Nicky Williams for many fruitful discussions, suggestions and advice.
Many thanks to the anonymous referees for their helpful comments.
%\end{small}
%% %% %% %, and the Fraunhofer for providing the examples from
%% %% %% %\cite{ACSLbyExample}. 
%% %% %% Special thanks to 
%% %% %% %Patrick Baudin,
%% %% %% Fran\c{c}ois Bobot, 
%% %% %% %Bernard Botella, 
%% %% %% Lo\"{i}c Correnson, 
%% %% %% %Pascal Cuoq, 
%% %% %% %Zaynah Dargaye, 
%% %% %% %Mathieu Lemerre, 
%% %% %% %Bruno Marre, 
%% %% %% %Virgile Prevosto, 
%% %% %% %Muriel Roger, 
%% %% %% Julien Signoles
%% %% %% and
%% %% %% Nicky Williams 
%% %% %% %and 
%% %% %% %Boris Yakobowski 
%% %% %% for many fruitful discussions, suggestions and advice.
%% %% %% %Many thanks to the anonymous referees for their helpful comments.

\bibliographystyle{splncs03}
\bibliography{biblio}

\end{document}